\documentclass{aa} \usepackage{graphics} \begin{document} 
\thesaurus{11(11.05.1; 11.09.1; 11.19.4} 
\title
{The globular cluster system around the low-luminosity S0 galaxy NGC 7457}

\author {S. Chapelon\inst{1}, V. Buat\inst{1,2}, D. Burgarella\inst{1}, M.
Kissler-Patig \inst{3}\thanks{Feodor Lynen Fellow of the Alexander von 
Humboldt Foundation}}
\institute {IGRAP, Laboratoire d'Astronomie Spatiale du CNRS, BP 8, 13376 
Marseille Cedex 12, France \and Laboratoire des interactions photons-mati\`ere, 
Facult\'e des Sciences de Saint J\'er\^ome, 13397 Marseille Cedex 20, France\ 
and European Southern Observatory, Karl-Schwarzschild-Str. 2, 85748 Garching, 
Germany}
\date{Accepted...}
\offprints{V. Buat}
\mail{buat@astrsp-mrs.fr}
\titlerunning{The globular cluster system of NGC 7457}
\authorrunning{Chapelon et al.}
\maketitle 
\begin{abstract} 

We investigate the globular cluster system around the low-luminosity S0 galaxy
NGC 7457 with deep B, V, I photometry. From the V photometry the total number
of globular clusters around the galaxy is estimated to be $\rm 178\pm 75$. The 
specific frequency is estimated to be $\rm S_N = 2.7\pm 1.1$ adopting $\rm M_V = 
-19.55$ mag for the galaxy.

We select 89 globular cluster candidates on the basis of their colors (B-I, V-I 
and B-V). The B-I and V-I color distributions are found  unimodal with a mean 
metallicity $\rm [Fe/H]\sim -1$ dex. Nevertheless the width of the B-I color 
distribution is found larger than that expected for a single population of 
globular clusters and is consistent  with the presence of more than one 
population. If the latter is the case, the two peaks would not be detected in 
B-I because of the small statistics and/or a small difference in the mean 
metallicity of the two populations. No metal poor globular clusters similar to 
those of the Milky Way halo are detected around NGC 7457.
 
\keywords {Galaxies: elliptical and lenticular, cD--Galaxies: individual: NGC 
7457--Galaxies: star clusters} 
\end{abstract}

\section{Introduction}

The globular clusters (GCs) are among the oldest known objects in the Universe
and can be considered as fossils of the formation of galaxies. GCs are simple 
coeval stellar systems which formed on a very short timescale during phases of 
intense star formation in galaxies. They are therefore more easily understood 
than mixed stellar field populations in a galaxy. The presence of globular 
cluster systems (GCSs) around most of the galaxies as well as the well known 
correlations found between some of their characteristics and those of the 
parent galaxies suggest that the formation of a galaxy and of its GCS are 
closely related (e.g. Harris \cite{harris}, Djorgovski \& Santiago 
\cite{djorgovski}).  

The alternative scenarios for the formation of early-type galaxies can be 
divided into two classes: the classical, long-lived scenario of a monolithic 
collapse which occurred in the early times of the Universe (e.g. Larson 
\cite{larson}) and the scenario issued from the hierarchical models of galaxy 
formation where the galaxies are formed through galaxy mergers (Kauffmann 
\cite{kauffmann}, Baugh et al. \cite{baugh}). During the unique collapse or for 
each major merging involving gas rich components, an intense star formation is 
expected with the likely formation of GCs. Indeed, it has been verified by the 
discovery of numerous proto-globular clusters by the Hubble Space Telescope in 
recently merging galaxies (and starbursting galaxies) in addition to the old 
ones from the progenitors (Schweizer \cite{schweizer} for a review): it 
demonstrates that GCs trace the major events in the star formation of the 
parent galaxy.

The consequences of the latter scenario on the GCSs have been extensively 
investigated by Ashman and Zepf (1992). To date the case of the {\it brightest} 
elliptical galaxies located in rich clusters is by far the best studied. A 
simple monolithic formation of these galaxies seems now to be ruled out by the 
observation of several sub-populations of GCs around these objects, indicating 
a more complex formation (e.g. Geisler et al. \cite{geisler}, Forbes et al. 
\cite{forbes}). Traces of merging is very frequent in these systems and merging 
has probably played a role in the galaxy formation and/or evolution. Recent 
spectroscopic studies of GCs around NGC 1399 (Kissler-Patig et al. 
\cite{kissbrod}) and M 87 (Cohen et al. \cite{cohen}) show that they probably 
formed massively in the early phases of the formation of these galaxies with a 
small contribution of recent events. However the case of cluster ellipticals may 
well not be appropriate to disentangle the models of galaxy formation since the 
hierarchical models also predict that the cluster ellipticals formed at 
relatively high redshift (e.g. Kauffmann \cite{kauffmann}).

In fact the case of field galaxies seems more interesting since the predictions 
of each model differ significantly for this class of objects. The hierarchical 
models predict their formation in recent dissipational mergers of gas-rich 
structures (e.g. Baugh et al. \cite{baugh}) which differ a lot from an early 
unique collapse. Early-type field galaxies with intermediate or low-luminosity 
are even more interesting. They might be the product of disk galaxies smaller 
than those involved in the formation of bright ellipticals (Kauffmann \& 
Charlot \cite{kauffcharl}). Indeed they generally exhibit a disky structure 
which is consistent with a formation in a dissipational merger of gas rich 
systems (Faber et al. \cite{faber}, Bender \cite{bender}) nevertheless the 
presence of gas during the merging is crucial to form new stars.

Unfortunately, the low-luminosity isolated early-type galaxies are by far less
studied than the bright cluster ellipticals. In general they show no obvious
hint of a past merger event (Kissler-Patig \cite{kissler}). Until now, the few
photometric studies of the GCSs around low-luminosity early-type galaxies are 
compatible with a single population and no complex formation (Kissler-Patig et 
al. \cite{kissfor}). But in most cases the low sensitivity of the colors used 
to estimate parameters like metallicity or age prevents a definitive 
conclusion.

In this paper we present the study of the GCS of the isolated, low-luminosity 
S0 galaxy NGC 7457. It is a disky galaxy (Michard \& Marchal \cite{michard}) 
and  its central light distribution is fitted by a power-law without evidence 
for the presence of a core (Lauer et al. \cite{lauer}). This galaxy does not 
show any trace of past interactions or merging (Schweizer \& Seitzer 
\cite{schweizseitz}). Nevertheless a steep central power-law in a disky galaxy 
can be the consequence of dissipative merging of gaseous galaxies (Bender 
\cite{bender}). Therefore the GCS around NGC 7457 may well have kept some 
traces of some merger events. It is in this context that we have undertaken the 
study of GCS around this galaxy.

The main characteristics of NGC 7457 are presented in table 1. Deep B, V and I
photometry will allow us to investigate the luminosity function of the GCS 
around this galaxy, to estimate the total number of GCs and the specific 
frequency and to study the color distribution of the GCs.

\begin{table*} 
\caption[]{General data for the target galaxy, NGC 7457 from the LEDA database 
(http://www-obs.univ-lyon1.fr/leda/). The B and V magnitudes are the total 
magnitudes $\rm B_T$ and $\rm V_T$ from the LEDA database corrected by us for 
Galactic extinction with $\rm A_G(B)=4.1\times E(B-V)$ and $\rm 
A_G(V)=3.1\times 
E(B-V)$ and E(B-V)=0.0525 from Burstein and Heiles (\cite{burstein}). V is the 
heliocentric velocity. The distance modulus is calculated with the heliocentric 
velocity corrected for the Local Group infall onto the Virgo cluster and 
assuming $\rm H_0 = 75$km/s/Mpc (Paturel et al. \cite{paturel}).} 
\begin{flushleft} 
\begin{tabular}{lllllllllll}
\hline 
Name & RA(2000) & DEC(2000) & {\sl l} & {\sl b} & Type & $\rm D_{25}$(arcmin) 
& V(km/s) & $\rm B$ & $\rm V$ & distance modulus \\ 
\hline
NGC 7457 & 23 00 59.9 & 30 08 39 & 96.22 & -26.9& E-SO & 3.98 
& 813 & 11.87 & 11.04 & 30.59\\ 
\hline 
\end{tabular} 
\end{flushleft} 
\end{table*}

\section{Observations and data reduction}

\subsection{Description of the data}

The B, V and I images were obtained at CFHT with the MOS instrument. A summary
of the exposures is given in table 2. The size of the field is $\rm 10\times 
10~ arcmin^{2}$ with a pixel size of 0.44$\rm\arcsec$. The equivalent 
exposure times are 10800 seconds in B, 4800 seconds in V and 2400 seconds in I 
with a seeing of 1.25$\rm\arcsec$, 1.04$\rm\arcsec$ and 1.11 
$\rm\arcsec$ respectively.

\begin{table} 
\caption []{Summary of the observations of the globular cluster system of NGC 
7457 at CFHT with the MOS instrument.}
\begin{flushleft} 
\begin{tabular}{llll} 
\hline 
Date & Filter & Exposure time & Number of exposures\\ 
\hline 
09/04/97 & B & 1200s& 5 \\
09/04/97 & I & 300s & 3 \\ 
09/04/97 & V & 600s& 4 \\
09/05/97 & I & 300s & 4 \\
09/05/97 & V & 600s& 4 \\ 
09/05/97 & B & 1200s & 4 \\ 
\hline  
\end{tabular} 
\end{flushleft} 
\end{table}

After having corrected all the images from bias and flat field, we combined the
frames for each filter using a median algorithm to remove the cosmic rays and
reduce the background noise. The galaxy profile has been next subtracted to
perform a more reliable photometry near the central parts of the galaxy. The
profile has been flattened by an iterative median-filter operation with a $\rm 
15\times 15$ pixels window. This smoothed frame was subtracted from the
original picture. Figure 1 is the final V picture on which photometry has been
performed.

\begin{figure}
\resizebox{\hsize}{!}{\includegraphics{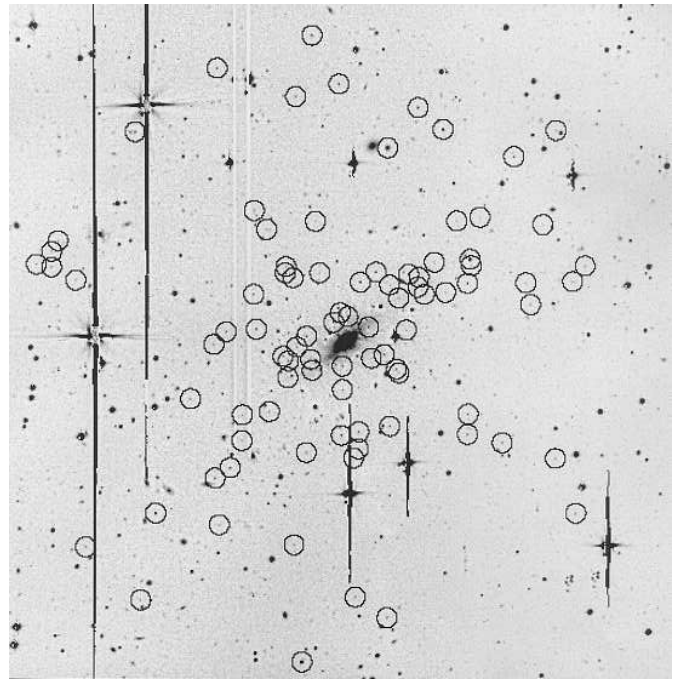}}
\vspace{2cm}
\caption[]{The globular cluster system around NGC 7457 in the V band. The 
galaxy profile has been subtracted. The 89 globular cluster candidates are 
marked with a circle. North is up and East on the left of the figure.}
\end{figure}
 
The detection has been performed in the V band since the observations are far
deeper than in the I band and somewhat deeper than in the B band.  This has been 
done using the DAOPHOT software in IRAF. 846 objects have been detected at a 
$\rm 6\sigma$ level above the local background. Such a high level of detection 
has been chosen since we wanted to keep only objects with accurate photometry 
measurements. A standard selection has been applied on the parameters 
"sharpness" and "roundness" given by the software. Finally we discarded some 
objects by visual inspection.

The detection in the B and I frames was made by searching by position for 
objects matching the V detections. Photometry was done in the three bands with
the PHOT software and only measurements with a photometric error lower than 0.2 
mag have been considered. From the 846 objects detected in V, 844 have a V 
magnitude ($\rm <\sigma (V)>\ =0.04$ mag), 763 have been measured in I ($\rm 
<\sigma (I)>\ =0.03$ mag) and 698 in B ($\rm <\sigma (B)>\ =0.07$ mag).

\subsection{The photometric zero-point}

The calibration in the three bands was done using 2 fields (including 11 
standard stars) with photometric standards (SA 110 and SA 113, Landolt 
\cite{landolt}). We determined the photometric zero-point by performing 
aperture photometry. The error in magnitude is estimated to be 0.02 mag in V, 
0.05 mag in I and 0.05 mag in B.

A Galactic color excess E(B-V)=0.0525 has been taken in the direction of NGC 
7457 from Burstein \& Heiles (\cite{burstein}) and each magnitude has been 
corrected using the values $\rm A_B=0.22$, $\rm A_V=0.16$ and $\rm A_I = 0.10$ 
mag calculated using the Galactic reddening curve of Pei (\cite{pei}).

\subsection{The completeness}

The knowledge of the completeness is essential to derive a density profile and
the total number of GCs around the galaxy. For each filter, we computed this 
completeness by adding artificial stars with the ADDSTAR routine of the DAOPHOT 
software. The magnitude was changed per bins of 0.25 mag or 0.5 mag. For each 
magnitude, 500 extra stars were added (over five runs in order to avoid 
crowding). The same procedure of detection as presented in section 2.1 has been 
applied. The 50\% completeness limit of a $\rm 6 \sigma$ detection is found to 
be 23.95 mag in V, 22.45 mag in I and 23.9 mag in B. The results are summarized 
in table 3.

\begin{table} 
\caption[]{Completeness percentages for the B, V and I observations as a 
function of the magnitude of the point sources.}
\begin{flushleft} 
\begin{tabular}{llll} 
\hline 
Magnitude & V band & I band & B band\\ 
\hline 
21.0 & & 94.2 & \\
21.5 & & 95.6 & \\
22.0 & & 92.6 & \\ 
22.25 & & 81.6 & \\
22.5 & 95.2 & 34.0 & \\
23.0 & 93.8 & & 94.8 \\ 
23.5 & 92.2 & & 91.6 \\ 
23.75 & 85.8 & & 78.6 \\
24.0 & 37.4 & & 35.2 \\ 
\hline  
\end{tabular}
\end{flushleft} 
\end{table}

\section{The selection of globular clusters candidates}

\begin{figure}
\resizebox{\hsize}{!}{\includegraphics{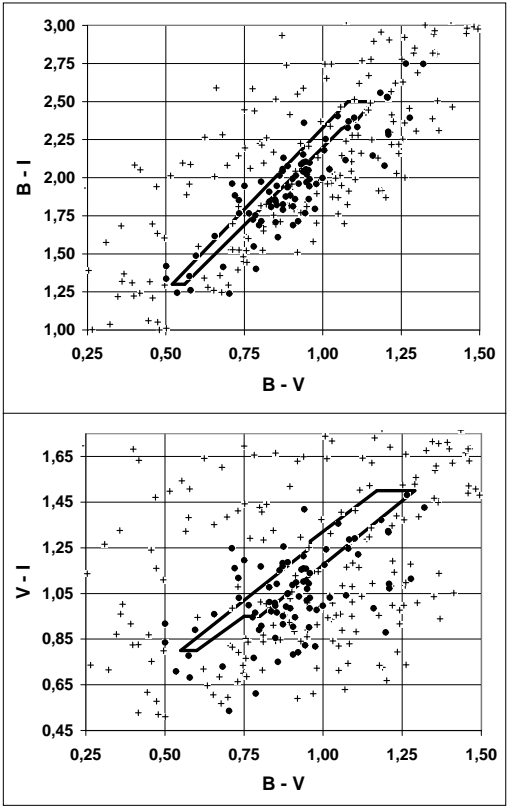}} 
\caption[]{Color-color plots for the globular cluster candidates around NGC 
7457. The point sources objects detected in the three photometric bands are 
reported with crosses. The typical internal photometric error as given by 
DAOPHOT is 0.3 mag for each color ($\rm 2\sigma$). The solid line is the 
envelope found by plotting synthesis models of stellar populations (Worthey 
\cite{worthey}) for a large range of metallicity (0.002 $\rm Z\odot$ to 2.5 
$\rm Z\odot$) and age (1 Gyr to 18 Gyr). The objects whose colors are 
compatible with the models when accounting for the photometric error given for 
each measurement by DAOPHOT are marked with a filled circle. We keep only the 
objects selected in both color-color diagrams.} 
\end{figure} 

The GCs are selected on the basis of their expected range in magnitude and on 
their colors. Therefore, to perform such a selection, a good photometry 
in at least two bands is necessary.  

We start from the list of point-like objects detected in the V frame and 
for which the V and I magnitudes are measured with a photometric error lower 
than 0.2 mag (see section 2.1). We exclude the central region of the galaxy 
(galactocentric distance lower than 15 arcsec) as well as the very external 
part of the image (galactocentric distance larger than 226 arcsec) in order to 
avoid too large a background contamination (see section 4.1 and figure 3). We 
are left with 412 objects.

First of all, we make a selection in magnitude, keeping only objects fainter 
than $\rm V>19$ mag. This corresponds to $\rm 3\sigma$ above the peak of the 
globular cluster luminosity function (GCLF) at the distance of NGC 7457 
(distance modulus of 30.59, see table 1), for a standard GCLF with $\rm 
M_V=-7.4$ mag and $\rm \sigma (M_V)=1.2$ mag (Harris \cite{harris}).

The selections in colors are made by comparing the V-I, B-V and B-I colors with
the predictions of stellar population synthesis models. In figures 2a and b are 
plotted these colors together with the envelope of points generated from the
models of Worthey (\cite{worthey}) for single stellar populations with a large
range of parameters (a metallicity ranging from $\rm 0.002~ Z\odot$ to $\rm 
2.5~Z\odot$ for ages comprised between 1 Gyr and 18 Gyr and three different 
initial mass functions (Salpeter (exponent x=-2.35 from $\rm M_l=0.21~ M\odot$ 
to $\rm M_u = 10~ M\odot$), Miller-Scalo (Miller \& Scalo, \cite{miller}, with 
$\rm M_l=0.1~ M\odot$ to $\rm M_u = 10~ M\odot$) and with an exponent x=-1.35 
(from $\rm M_l=0.33~ M\odot$ to $\rm M_u = 10~ M\odot$)). We select all objects 
whose colors are consistent with the envelope of the models adopting a $\rm 
2\sigma$ uncertainty on each color given by the DAOPHOT software. We use the 
models of Bertelli et al. (\cite{bertelli}): they lead to a similar selection 
of GC candidates. The resulting range of acceptable colors is $\rm 
0.80<V-I<1.50,~ 1.3<B-I<2.7,~ 0.5<B-V<1.6$ but our selection is more refined 
than adopting simple color ranges since we check the compatibility of the three 
colors by their locus in color-color diagrams. For the objects which are not 
detected in B (5\% of the sample) a range of plausible values for their B 
magnitude is estimated from the models and their V-I colors. Then we reject the 
objects which must be detected in B (the limiting B magnitude is taken at 23.5 
mag (90\% of completeness)). After these selections we are left with 89 GC 
candidates, out of which only 3 have no B measurement.

\section{The number of globular clusters and the specific frequency}

\subsection{The density profile}

\begin{figure}
\resizebox{\hsize}{!}{\includegraphics{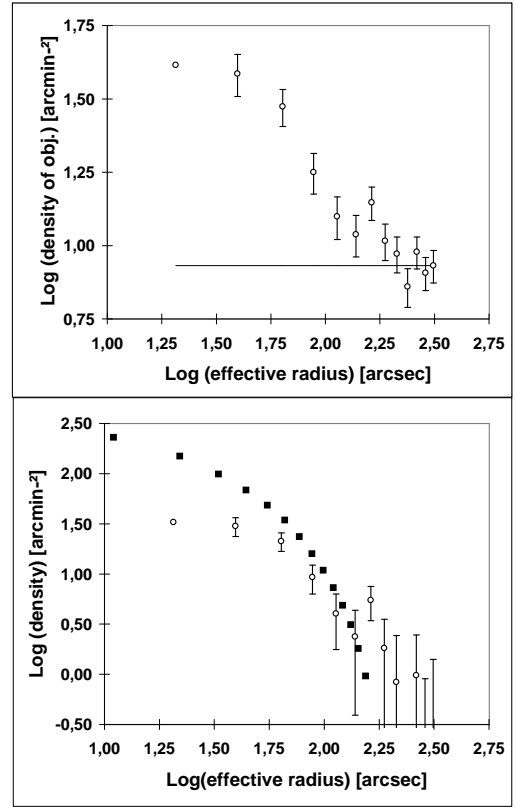}}
\caption[]{Upper panel: the surface density profile (open circles) of 
point-like objects detected around NGC 7457 in the V frame from the values 
quoted in table 4. The profile flattens at about 225 $\rm\arcsec$ to a constant 
level of $\rm 8.5\pm 0.9~ objects/arcmin^{2}$ (line). Lower panel: the surface 
density profile of the globular cluster system (open circles) compared with the 
isophotal galaxy light in V plotted in arbitrary units with filled squares.}
\end{figure}

We study the density profile of GCs using the {\it unselected} V detections 
down to V=23.9 mag, our 80\% completeness limit. To compute this density profile 
we bin the sample in elliptical rings having the same inclination and 
ellipticity as NGC 7457 ($\rm \epsilon = 0.41$ and $\rm p.a. = 130\degr$, LEDA 
database). The major axis step is about 25 $\rm\arcsec$ and the minor 15 
$\rm\arcsec$. The counts are then corrected for magnitude incompleteness. No 
correction for geometrical incompleteness is necessary except for the last ring 
(major axis from 300 to 325 $\rm\arcsec$ from the center) for which we lack 10\% 
of the surface because of the limited size of the observed field. We do not 
consider the most central part of the galaxy which is saturated (for radii 
smaller than 15 $\rm\arcsec$). Inside the geometrical bins, we group the data in 
magnitude bins of 0.5 mag to correct for incompleteness at different magnitudes. 
The counts are shown in table 4. For the first ring, we could not estimate the 
Poissonian error because we are too close from the center of the galaxy.

\begin{table*} 
\caption[]{The globular cluster counts around NGC 7457. The counts are 
performed in elliptical rings down to a magnitude $\rm V =23.9$ mag. In column 
2 are reported the counts without any correction. In column 4 and 5 the counts 
are corrected for completeness, the quoted errors are Poissonian. In column 6 
the density of GCs is  corrected for the background contamination, the quoted 
error accounts for the error on the counts (column 4) and on the background ( 
$\rm 8.5\pm 0.9~ objects/arcmin^{2}$, see text).}
\begin{flushleft} 
\begin{tabular}{llllll} 
\hline 
Ring major axis & Objects counts & Ring surface & Corrected counts & Surface 
density & Density profile\\ 
arcsec & & arcmin$^{2}$ &  & objects/armin$^{2}$ & GCs/armin$^{2}$\\ 
\hline
15-25 & 8 & 0.21 & 8.5 $\pm$ ? & 41.3 $\pm$ ? & 32.7 $\pm$ ?\\ 
25-50 & 34 & 0.97 & 37.2 $\pm$ 6.1 & 38.5 $\pm$ 6.3 & 30.0 $\pm$ 6.4\\ 
50-75 & 44 & 1.61 & 47.9 $\pm$ 6.9 & 29.8 $\pm$ 4.3 & 21.2 $\pm$ 4.4\\
75-100 & 37 & 2.25 & 40.1 $\pm$ 6.3 & 17.8 $\pm$ 2.8 & 9.3 $\pm$ 3.0\\ 
100-125 & 34 & 2.90 & 36.4 $\pm$ 6.0 & 12.6 $\pm$ 2.1 & 4.0 $\pm$ 2.3\\ 
125-150 & 36 & 3.54 & 38.6 $\pm$ 6.2 & 10.9 $\pm$ 1.8 & 2.4 $\pm$ 2.0\\ 
150-175 & 54 & 4.18 & 58.6 $\pm$ 7.7 & 14.0 $\pm$ 1.8 & 5.5 $\pm$ 2.0\\
175-200 & 46 & 4.83 & 50.0 $\pm$ 7.1 & 10.4 $\pm$ 1.5 & 1.8 $\pm$ 1.7\\
200-225 & 48 & 5.47 & 51.3 $\pm$ 7.2 & 9.4 $\pm$ 1.3 & 0.8$\pm$ 1.6\\
225-250 & 41 & 6.11 & 44.3 $\pm$ 6.7 & 7.2 $\pm$ 1.1 & -1.3 $\pm$ 1.4\\ 
250-275 & 60 & 6.76 & 64.3 $\pm$ 8.0 & 9.5 $\pm$ 1.2 & 1.0 $\pm$ 1.5\\ 
275-300 & 56 & 7.40 & 59.7 $\pm$ 7.2 & 8.1 $\pm$ 1.0 & -0.5$\pm$ 1.4\\
300-325 & 58 & 7.27 & 62.1 $\pm$ 7.9 & 8.5 $\pm$ 1.2 & 0.0$\pm$ 1.4\\ 
\hline   
\end{tabular} 
\end{flushleft} 
\end{table*}
In figure 3 are plotted the radial variation of the surface density, 
distribution for our point-like V detections together with the galactic V light 
profile. The background contamination can been estimated when the density 
profile reaches a constant value. Practically this occurs for galactocentric 
distances larger than 200 arcsec. The average of the counts in the five last 
rings (table 4, column 5) gives $\rm 8.5\pm 0.9~ objects/arcmin^{2}$ in 
agreement with the deep counts down to $\rm V = 23.9$ mag, our 80\% limit 
completeness (Peterson et al. \cite{peterson}, Tyson \cite{tyson}). After 
applying corrections for incompleteness, the total number of objects down to 
this magnitude is estimated to be $\rm 537\pm 38$. The background is subtracted 
statistically over an area of $\rm 46.34~arcmin^{2}$ (thus $\rm 394\pm 46$ 
objects) and we are left with a total of $\rm 143\pm 60$ GC candidates.

The density profile of the GCs follows the galaxy light profile. The fit of a 
power-law on the profiles ($\rm \propto r^{-\alpha}$) gives $\rm \alpha=2.10\pm 
0.26$ for the GC density profile and $\rm \alpha=2.02\pm 0.23$ for the galaxy 
light profile. We used all the data points except for the first bin for which 
we do not estimate the error. Such a slope ($\rm \alpha\simeq -2$) for the 
GC profile of NGC 7457 whose absolute V magnitude is -19.55 mag is in agreement 
with the correlation found between this slope and the absolute magnitude of the 
parent galaxy, fainter galaxies having a steeper distribution of GCs (e.g. 
Harris \cite{harris}, Kissler-Patig \cite{kissler}). A similar profile for the 
galaxy light and the GCS seems not to be the rule for bright ellipticals (e.g. 
Ashman \& Zepf \cite{ashzepf}) but the case of low-luminosity early-type 
galaxies is less clear since very few data are available for them.

\subsection{The globular cluster luminosity function}

\begin{figure}
\vspace{2cm}
\resizebox{\hsize}{!}{\includegraphics{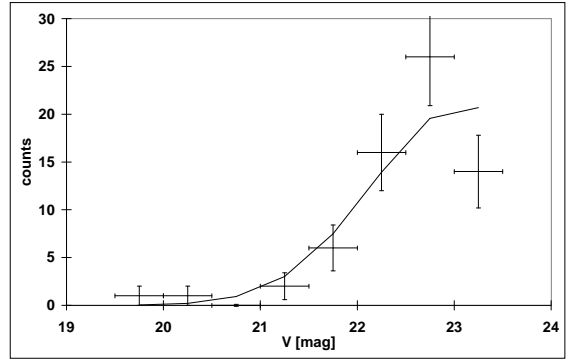}}
\caption{The globular cluster luminosity distribution in the V band fitted by a 
gaussian function with mean and $\rm \sigma$ equal respectively to 23.1 mag and 
0.9 mag.}
\end{figure}

To compute the globular cluster luminosity function we would need to consider 
only genuine GCs. To this aim, we use the GC candidates selected in section 3 
from V and I detections.

We have to account for incompleteness effects due to the magnitude limits in V
and I. The range of V-I colors for the GC candidates is estimated from the 
models of Worthey namely $\rm 0.80<V-I<1.5$. Since the GC candidates are first 
detected in the V band, we apply a selection which ensures the full range of 
possible V-I color for each detected candidate. With a completeness limit 
(50\%) at 22.45 mag for I and 23.95 mag for V we are limited by the I 
detections and we have to truncate the sample at V=23.25 mag which is reduced 
to 65 GC candidates.

The GCLF is shown for V$<$ 23.25 mag in figure 4. A gaussian distribution has 
been fitted to the bins not affected by incompleteness. The parameters of the 
gaussian are $\rm M_V=23.1\pm 0.2$ mag for the mean and a dispersion of $\rm 
\sigma_V=0.9\pm 0.3$ mag. The mean value that we find is fully consistent with 
a mean absolute magnitude $\rm M_V=-7.4$ mag for the distance modulus of NGC 
7457 (table 1). The dispersion we find for the GCLF of NGC 7457 is marginally 
consistent with those found in galaxies like M 31 or the Milky Way with $\rm 
\sigma_V=1.2\pm 0.1$ mag whereas for bright ellipticals the dispersion seems 
larger with $\rm \sigma_V \simeq 1.4$ mag (e.g. Ashman \& Zepf \cite{ashzepf}).

\subsection{The specific frequency}

The number of GCs in an early-type galaxy has been found to scale with the 
parent galaxy luminosity (e.g Djorgovski \& Santiago 1992). Therefore following 
Harris \& van den Bergh (1981) the GC counts are  normalized per 
absolute visual magnitude of $\rm M_V = 15$ mag and a specific frequency is 
defined as $\rm S_N = N\times 10^{0.4\cdot (M_V~+~15)}$ where N is the total 
number of GCs in the galaxy and $\rm M_V$ its absolute visual magnitude. $\rm 
S_N$ is a measure of the GC efficiency. $\rm S_N$ is found to be $\rm \sim 5$ 
for most early-type galaxies with $\rm -19 > M_V > -22$ although with a large 
dispersion. Brighter early-type galaxies exhibit in average a larger $\rm S_N$, 
some giant ellipticals in rich clusters having a very large specific frequency 
$\rm S_N>10 $ (Kissler-Patig \cite{kissler}).

In the section 4.1 we estimated the total number of GCs down to $\rm M_V = 
23.9$ mag. Then integrating over the entire luminosity distribution of the 
section 4.2 we deduce a total number of $\rm 178\pm 75$ clusters. This 
translates to a specific frequency $\rm S_N = 2.7\pm 1.1$ assuming an absolute 
V magnitude $\rm M_V =-19.55$ mag for NGC 7457. We can compute in the same way 
the specific frequency from the number of detected GC down to V = 23.1 mag (the 
turn-over of the GCLF) and, after multiplying by two, we are left with $132 \pm 
58$ objects thus $\rm S_N = 2.0\pm 0.9$.

\section{The color distributions of the globular clusters}

\begin{figure} 
\resizebox{\hsize}{!}{\includegraphics{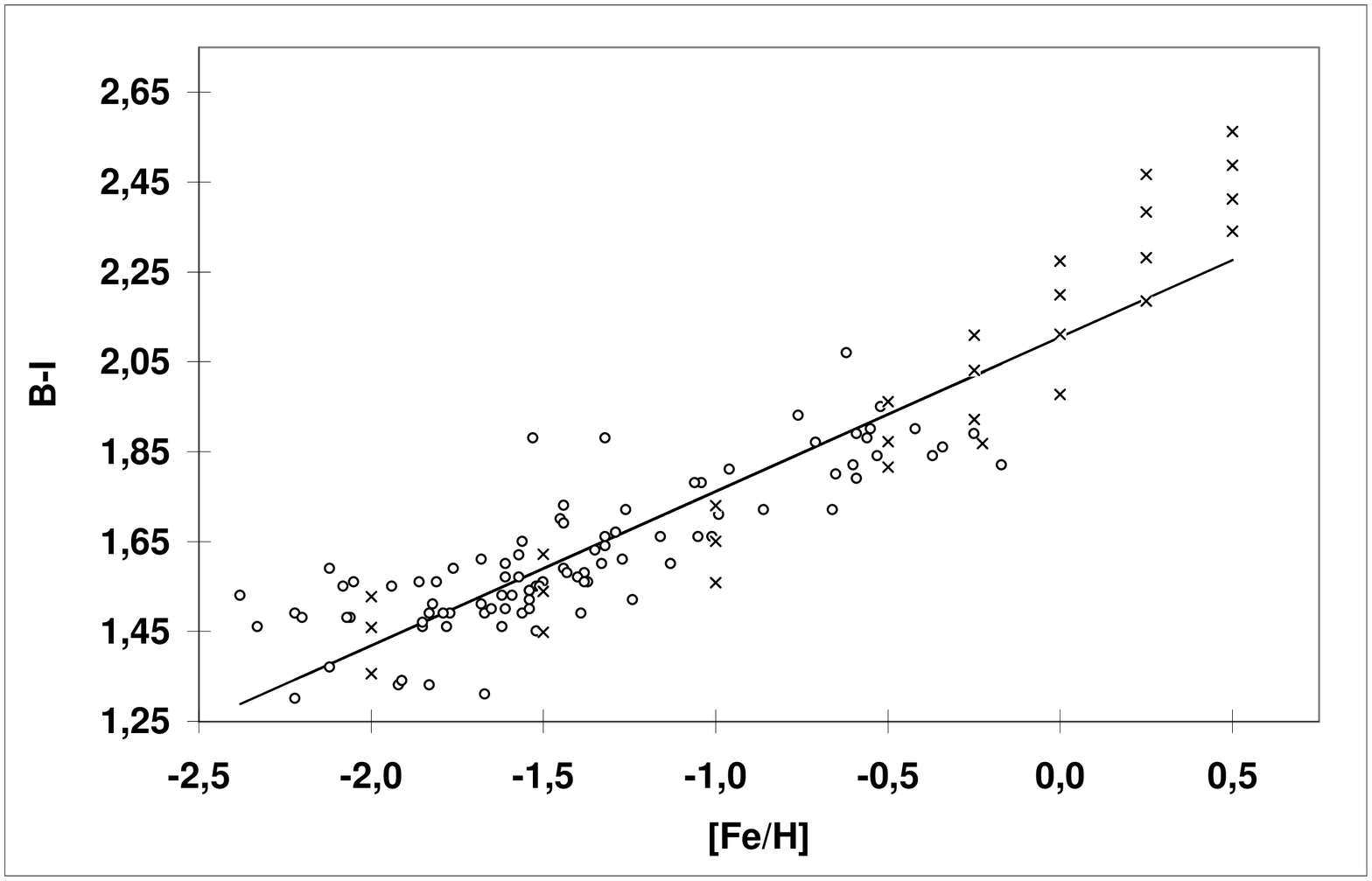}} 
\caption[]{B-I color versus metallicity of the MW globular clusters (open 
circles) together with the the results of Worthey's models for old homogeneous 
stellar populations (crosses). The solid line is the result of the linear 
regression.}
\end{figure}

Now we study the V-I and B-I color distributions of the GCs around NGC 7457 in 
order to search, if any, the presence of sub-populations with different 
metallicities. Indeed the B-I color and in a lesser extent the V-I color are 
sensitive to the metallicity provided that the GCs are old ($> 5$ Gyr). In case 
of old stellar populations these two colors are poor tracers of the age of the 
stellar populations (Worthey \cite{worthey}, Kissler-Patig et al. 
\cite{kissbrod}).

First we need to calibrate the V-I and B-I colors as a function of the
metallicity. Empirical relations based on the observation of the Milky Way (MW)
GCs are commonly used (e.g. Couture et al. \cite{couture}) but these relations 
become very insecure outside the metallicity range of the MW GCs ($\rm -2.0\le 
[Fe/H] \le -0.5$). Kissler-Patig et al. (1998) have extended the comparison 
between V-I and the metallicity deduced from spectroscopic data to higher 
metallicities thanks to their observations of NGC 1399 and found a flatter 
relation of $\rm [Fe/H]$ as a function of V-I than when only the MW GCs are 
considered. In this paper, we will adopt the relation of Kissler-Patig et al. 
(\cite{kissbrod}):
$$\rm [Fe/H] = (-4.50\pm 0.30) + (3.27\pm 0.32) (V-I)$$

For the relation between the B-I color and the metallicity of GCs we re-analyse 
the problem by using the data on the MW clusters (Harris 1996, McMaster 
database) together with the synthesis models of Worthey (1994) for old stellar 
populations ($> 5$ Gyr) with a large range of metallicities ($\rm -2.0\le 
[Fe/H] \le 0.5$). We consider only old ages for the models in order to be 
consistent with the age of the MW GCs. Therefore, the calibration relation will 
be valid for old stellar populations only ($> 5$ Gyr). When comparing the 
colors listed in Worthey (1994) with the observed MW GCs, a small shift in 
color is needed. Assuming an average age of 12 Gyr for the MW GCs, we get a 
slight correction (-0.09 in V-I, -0.14 in B-I) to apply to the models. These 
corrections are in reasonable agreement with the color shifts suggested by 
Worthey. The data are presented in figure 5. A linear regression gives:

$$\rm [Fe/H] = (-5.34\pm 0.42) + (2.44 \pm 0.30) (B-I)$$

Given the uncertainties about these calibration relations, we have checked that 
our results do not depend on the exact choice of the conversion formulae. Indeed 
we have used different relations like those of Couture et al. for both V-I and 
B-I as well as the relation we find between [Fe/H] and V-I following the same 
method as described in the last paragraph (MW data and synthesis models).

\subsection{The V-I color distribution}
\begin{figure}
\resizebox{\hsize}{!}{\includegraphics{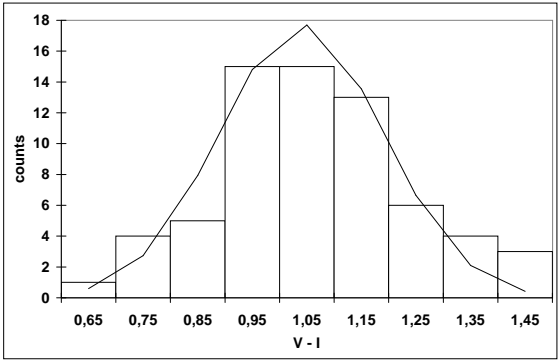}}
\caption[]{The histogram of the V-I color distribution for the globular 
clusters of NGC 7457. The typical internal photometric error is 0.10 mag as the 
size of the bins.}
\end{figure}

The V-I color histogram is reported in figure 6 for the sample of 65 GCs with 
$\rm V<23.25$ mag (see section 4.2). The distribution is found unimodal using 
the KMM mixture-modeling algorithm (Ashman et al. \cite{ashbird}). The gaussian 
distribution fitted on the data has a mean $\rm <V-I>\ =1.04\pm 0.02$ mag and a 
standard deviation $\rm \sigma(V-I)=0.15\pm 0.04$ mag. We can also estimate the 
mean and standard deviation using maximum likelihood estimators (Pryor \& 
Meylan \cite{pryor}), this method has the advantage of being performed without 
any hypothesis on the shape of the distribution. We find very similar results 
i.e. $\rm <V-I>=1.06 \pm 0.02$ mag and $\rm \sigma(V-I)=\ 0.16\pm 0.02$ mag. 
Therefore the observed standard deviation is only slightly larger than the mean 
internal photometric error for the sample ($\rm \sigma_{err}(V-I)=0.1$ mag).\\
A mean $\rm <V-I>\ =1.04$ mag translates to $\rm <[Fe/H]>\ =-1.1$ dex. With an 
absolute V magnitude of -19.55 for NGC 7457 such a mean is consistent with the 
correlation found between the absolute magnitude of the parent galaxy and the 
mean metallicity of the GCSs (e.g. Ashman \& Zepf \cite{ashzepf}).

\subsection{The B-I color distribution}
\begin{figure}
\resizebox{\hsize}{!}{\includegraphics{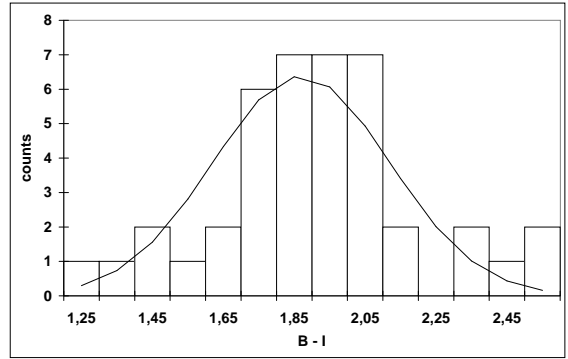}}
\caption[]{The histogram of the B-I color distribution for the globular 
clusters of NGC 7457. The typical internal photometric error is 0.1 mag of the 
order of the size of the bins (0.1 mag).}
\end{figure}

To compute the B-I color distribution we have to select GC candidates detected 
in V and with a measured magnitude in I and B. Given the deepness of the V 
detection as compared to the I and B ones, the resulted sample is equivalent to 
a selection based only on B and I detections. Once again we only keep objects 
with $\rm V>19$ mag.

We have also to be as complete as possible in the B and I bands for the 
analysis of the B-I color distribution. The range of B-I colors is estimated 
from the models of Worthey used for the selection of the GCs candidates i.e. 
$\rm 1.3<B-I<2.7$. As before, we adopt a completeness limit at 50\% which 
corresponds to 22.45 mag in I and 23.9 mag in B. To ensure the full range of 
possible colors for each candidate detected in B we must truncate the sample
at B=23.7 mag. The sample is reduced to 41 objects.

The B-I color histogram is reported in figure 7. The distribution is larger
than for the V-I color but it is also found unimodal with the KMM test. Such a
result is expected since Ashman et al. (\cite{ashbird}) have shown that a
bimodality cannot be easily detected with the KMM algorithm when the number of 
data is lower than 50 as it is the case here. The fit with a gaussian 
distribution gives a mean $\rm <B-I>\ =1.87\pm 0.02$ mag leading to a $\rm 
[Fe/H]=-0.8$ dex consistent with the mean metallicity deduced from the V-I 
distribution given the uncertainties on the calibration relations; the observed 
standard deviation is $\rm \sigma(B-I)=0.25\pm 0.06$ mag. Using maximum 
likelihood estimators we find $\rm <B-I>\ =1.91\pm 0.04$ mag and $\rm 
\sigma(B-I)=0.28\pm 0.03$ mag. Therefore the observed standard deviation of the 
B-I distribution is larger than the mean internal error $\rm 
\sigma_{err}(B-I)=0.1$ mag.

\subsection{A hidden bimodality?}

The V-I and B-I distributions of the GCs around NGC 7457 are found unimodal. 
Before discussing the implications of this result we discuss in what conditions 
a bimodality would be really detected. First, as already discussed it is 
imperative to detect more GCs and therefore to perform a deeper photometry in 
at least two bands. Moreover Kissler-Patig et al. (\cite{kissfor}) have 
outlined that, even if present, different populations of GCs are probably hard 
to discriminate in low-luminosity galaxies. A conspiracy between age and 
metallicity of the two populations could mimic a unimodal V-I distribution, the 
V-I color being very sensitive neither to age nor to metallicity.

Another difficulty is that even if the metallicity distribution of the GCS
is bimodal the two peaks might be too close to be disentangled. For example 
Forbes et al. (\cite{forbes}) have found a correlation between the mean 
metallicity of the metal-rich population of GCs and the luminosity of bright 
ellipticals whereas there is almost no correlation between the metallicity of 
the metal-poor GCs and the parent galaxy luminosity. If such a correlation also 
holds for fainter galaxies, their redder GC population will be less metal-rich 
and its color peak closer to that of their metal-poor GC population than it is 
the case in the bright ellipticals. Therefore, the two peaks in the metallicity 
distribution, if they exist, will be hardly detected in these galaxies.  

In order to investigate more quantitatively this effect, we can try to estimate
what sort of bimodality is detectable with our data. Ashman et al. 
(\cite{ashbird}) have tested the bimodal detectability as a function of $\rm
\Delta \mu$, the difference between the two means divided by the standard
deviation of the two sub-populations supposed to have the same dispersion.  For
a total number of $ \rm \sim 50$ GCs, a bimodality is detected as soon as $\rm
\Delta \mu \ge 3$. If we adopt a typical $\rm \sigma ([Fe/H])=0.3$ dex for
each sub-population we find $\rm \Delta [Fe/H]\ge 0.9$ dex. However, we must
note that this estimation is based on the assumption that the two
sub-populations have the same number of objects. As noted by Kissler-Patig et
al. (\cite{kissfor}, see also section 6.2), if there is not an equal number of
GCs in each sub-population, they cannot be separated when $\rm \Delta 
[Fe/H]\sim 1$ dex. $\rm \Delta [Fe/H]=0.9$ dex corresponds to $\rm \Delta 
(V-I)= 0.3$ and $\rm \Delta (B-I)= 0.4$ mag according to the calibration 
formulae given above. Thus, under these hypotheses, the two peaks might be 
separated.

A 1-dex separation seems to a rough order of magnitude of what we may find
in GCSs. Indeed, in bright ellipticals the difference between the two 
metallicity peaks approximately or slightly larger than 1 dex (e.g. Ashman \& 
Zepf \cite{ashzepf}). Also, in the Milky Way, a similar gap has been measured 
between the metallicity of the halo and disk/bulge GCs (Armandroff \& Zinn 
\cite{armandroff}). 

The bottomline is that with our data we can hope to separate two peaks of a 
bimodal distribution similar to those found around bright ellipticals or the 
Milky Way provided that the two sub-populations are roughly equally 
represented. Any bimodal distribution with two metallicity peaks separated by 
less than 1 dex could not be separated.

\section{Discussion}

NGC 7457 is a S0 low-luminosity field galaxy. S0 galaxies are thought
to be the continuation from disky ellipticals to early-type spirals in the
Hubble sequence (Kormendy \cite{kormendy}). We can expect that S0 and disky
ellipticals have been formed through similar processes. Therefore this type of
galaxy is particularly interesting for testing the scenarii of galaxy 
formation, in particular whether these galaxies formed in a dissipational 
merger event (see section 1).

The simple model of a spiral--spiral merger predicts several properties for the
GCS of the remnant galaxy (Ashman \& Zepf \cite{ashman}): the newly produced 
GCs should rise the specific frequency by a factor of two, the new clusters and 
stars will be more concentrated than the GCS of the progenitors, and the new 
clusters should appear as a second population in the color distribution.

At face value NGC 7457 does not seem to confirm the hypothesis of a formation by 
spiral--spiral merging, at least when compared to the predictions of the 
simple model of Ashman \& Zepf (\cite{ashman}). NGC 7457 is found to have a 
rather low number of GCs steeply distributed around the galaxy, and the color 
distribution of these GCs appears unimodal. Indeed, the specific frequency $\rm 
S_N=2.7\pm 1.1$ is compatible with the range observed in the four studied Sa 
and Sab galaxies (ranging from 0.7 to 3.5 with a mean of 2.0 and a dispersion 
of 1.0, taken from Ashman \& Zepf \cite{ashzepf}). The surface density profile 
of the GCs follows the one of the stars (see section 4.1). The absence of a 
recent merging event is suggested by the lack of any fine structure as defined 
by Schweizer \& Seitzer (\cite{schweizer}) for this galaxy.

However two results might constrain formation theories. First the color 
distributions of the GCs appear broader than for a single population such as, 
e.g., the Milky Way halo clusters. No clear bimodality has been detected. 
Nevertheless given the poor statistics it has been shown in section 5.3 that it 
does not exclude necessarily the presence of distinct sub-populations. And, 
second, the peak of the color distribution appears relatively red.

\subsection{A broad color distribution}

Does the unimodality of the color distribution imply the absence of two
distinct GC populations? In spite of the difficulties discussed in the section 
5.3, the narrow V-I distributions found in low-luminosity early-type galaxies 
(Kissler-Patig et al. \cite{kisskohl}), together with a narrow luminosity 
function, can be used to exclude large difference in both metallicity and age 
within each galaxy, and therefore a recent gas-rich merger (z$<$1) 
(Kissler-Patig et al. \cite{kissfor}). 

The availability of the B-I colors for the GCs of NGC 7457 allows us to go 
further in the analysis. Indeed the B-I color is roughly twice as sensitive to 
metallicity than the V-I color (Couture et al. \cite{couture}). But an 
intrinsic difficulty with low-luminosity galaxies is their small number of GCs. 
Bimodality was shown to be undetectable in a dataset containing less than 50 
objects (see section 5.3). Nevertheless, as we will see below, we can expect to 
observe some differences between the widths of a unimodal and a bimodal 
distributions. The dispersion in color of a ``single'' population of GCs can be 
estimated from the halo GCs of the MW. On the one hand, we can convert the 
metallicity dispersion into a color dispersion. With $\rm \sigma ([Fe/H])=0.3$ 
dex (Armandroff \& Zinn \cite{armandroff}) we expect a $\rm \sigma (V-I) \sim 
0.05$ mag and $\rm \sigma (B-I)=0.1$ mag from the calibration relations of 
section 5. On the other hand, we can measure the dispersion from the V-I and B-I 
data of the $\sim 80$ halo GCs in the McMaster catalog (Harris 1996), and obtain 
$\rm \sigma (V-I)=0.05\pm 0.01$ mag and $\rm \sigma (B-I)=0.09\pm 0.01$ mag in 
excellent agreement with the first values. In V-I, the genuine dispersion in 
metallicity is extremely difficult to derive from the V-I colors, the expected 
dispersion being lower than the typical photometric errors of 0.1 mag. In B-I, 
however, typical photometric errors and intrinsic dispersion of a single 
population are comparable, so that several populations would broaden the color 
distribution to a detectable level.

In NGC 7457 the dispersion in V-I is only slightly larger than the internal 
error (0.15 mag against 0.10 mag) and we estimate that the true dispersion is 
$\rm \sim 0.11\pm 0.11$ mag according to the relation $\rm \sigma^2_{obs} = 
\sigma^2_{err}+\sigma^2_{true}$. This standard deviation in V-I translates to a 
$\rm \sigma([Fe/H])\sim 0.4\pm 0.4$ dex when using the calibration relation 
given in the precedent section. The error is estimated by accounting for the 
uncertainties on the determination of the standard deviations and on the 
calibration formula given in section 5. The estimate is very insecure due to 
the large error on $\rm \sigma(V-I)$.\\
For the B-I color we find a dispersion (0.25 mag) clearly broader than the
combination of a single population and photometric errors. The above relation 
leads to $\rm \sigma_{true}=0.23 \pm 0.07$ mag or $\rm \sigma([Fe/H])=0.6\pm 
0.2$ dex. This value is compatible with the value tentatively deduced from the 
V-I distribution.

Such a dispersion in metallicity seems intermediate between the one of a single 
population and the total dispersion of the GC populations in bright 
ellipticals. Indeed single populations such as the Galactic halo GCs (Armandroff 
\& Zinn \cite{armandroff}), or the GCs around M 81 or M 31 (Perelmuter \& Racine 
\cite{perelmuter}) have $\rm \sigma ([Fe/H])\sim 0.3$ dex; the individual 
components of the bimodal distribution in NGC 4472 have similar dispersions of 
$\sim 0.38$ dex in $\rm [Fe/H]$ (Geisler et al. \cite{geisler}). In contrast, 
the total dispersion of the system in M 87 is $\rm \sigma([Fe/H])=0.65$ dex (Lee 
\& Geisler \cite{lee}), in NGC 4472 $\rm \sigma([Fe/H])=0.7$ dex (Geisler et al. 
\cite{geisler}).

Therefore, while the distribution in metallicity of the GCs around NGC 7457 is 
found to be unimodal, the width of the GC metallicity distribution is 
compatible with the presence of different populations probably less separated in 
metallicity than in the giant clusters ellipticals. This suggests a 
significantly different chemical enrichment of the GCs in NGC 7457 than, e.g., 
the halo population of the Galaxy.

\subsection{The mean colors of the globular clusters: implications on the   
scenarios of formation}

With $\rm M_V=-19.55$ and a mean metallicity of [Fe/H]$\ \simeq -1$ dex for its 
GCs, NGC 7457 follows the general trend found between the absolute magnitude 
of the galaxies (spirals $+$ ellipticals) and the [Fe/H] value of their GCs 
(e.g. Brodie \& Huchra \cite{brodie}, Ashman \& Zepf \cite{ashzepf}). 
Nevertheless, the spirals seem to have a lower GC metallicity as compared to 
ellipticals of similar luminosity and the mean metallicity of [Fe/H]$\ \simeq 
-1$ dex for the GCs around NGC 7457 is consistent with the  mean values found 
for the metallicities of the GCs around the bright elliptical galaxies ($\rm 
M_V\le -20$, e.g. Ashman \& Zepf \cite{ashzepf}).
 
We can also compare more quantitatively the color distribution of the GCs around 
NGC 7457 with that of the Galactic GCs. In addition to its broad dispersion, the 
mean B-I color found for the GCs of NGC 7457 (B-I$\ \simeq1.9$ mag, see section 
5.2) is comparable to the mean of the Galactic disc/bulge GCs (B-I$\ \simeq 1.9$ 
mag, as derived from the McMaster catalog).\\ 
Moreover, Monte Carlo simulations of B-I color distributions similar to ours 
show that any metal-poor (B-I=1.55 mag, the mean color of the metal-poor 
clusters in the Galaxy) population as large as 20\% to 30\% of the metal-rich 
(B-I$=1.92$ mag) one would be detected. Therefore we can conclude to the absence 
of any significant population of metal-poor clusters similar to that of the MW 
halo.

It is likely that such blue globular clusters were never present in NGC 7457
since NGC 7457 is an isolated galaxy ($\rm \rho=0.13~~ galaxies/Mpc^{3}$) and
shows no signs of any interaction.  Thus the loss of blue GCs loss through
stripping seems excluded.\\
The formation in a spiral--spiral merger would imply the presence of blue GCs 
from the progenitor spirals unless the latter did not host blue GCs like the MW. 
In situ formation models usually explain blue GCs as formed in the early stage 
of the galaxy. To fit the absence of blue clusters in such scenarios an early 
epoch of star formation (to enrich the gas) without any formation of GCs would 
be required.\\
We can also explore the possibility that the halo population of NGC 7457 and
therefore of its progenitors in case of merging would be slightly redder than
the MW one. For both the merger and accretion models, no variation of the color
of the blue cluster population as a function of the parent galaxy luminosity is
expected as it seems to be confirmed by Forbes et al (\cite{forbes}).  Indeed
their mean metal-poor peak $\rm <[Fe/H]>\ = -1.2\pm 0.3 ~$ dex obtained for
bright ellipticals translates to $\rm B-I=1.7$ and V-I = 1.0 (see also Neilsen 
et al. \cite{neilsen}). An extrapolation of the trend that Forbes et al.  found
between the metal-rich peak and the galaxy luminosity gives $\rm [Fe/H]\simeq
-0.6 $ dex and thus $\rm B-I\simeq 1.95$, $\rm V-I\simeq 1.2 $.  Such a bimodal
distribution is consistent with our data since the two peaks could not be
separated neither in V-I nor in B-I given the proximity of their colors (see
section 5.3).\\
Finally, the absence of blue GCs around NGC 7457 could be explained in a 
scenario as the one advanced by C\^ot\'e et al. (\cite{cote}) where a galaxy
forms GCs with a mean metallicity proportional to its luminosity and gains its
metal-poor GCs by accreting smaller galaxies surrounding it. In this case, the
absence of a significant number of metal-poor GCs in NGC 7457 could then be due
to a lack of dwarf galaxies around this isolated galaxy.

Two galaxies already were reported to lack blue GCs: NGC 3923 and NGC 3311. Zepf 
et al. (\cite{zepf}) have discussed various scenarios to explain the high mean 
metallicity ([Fe/H] = -0.56 dex) and the very few clusters with [Fe/H] $<$ -1 
dex around NGC 3923, a luminous field elliptical. They propose several 
explanations around the merger picture, in which the number and metallicity of 
the clusters formed during the merging as well as those of pre-existing clusters 
around the progenitors might vary. Finally more or less accretion of satellite 
galaxies and their metal-poor GCs can also modify the final color distribution 
of the system.\\
In case of NGC 3311, the central cD in the Hydra cluster, Secker et al. 
(\cite{secker}) attribute the almost complete lack of blue GCs (less than 10\% 
of the GCs appear to have $\rm [Fe/H]<-1$ dex) to different history of 
pre-enrichment of the giant galaxy in the frame of an in situ formation of 
clusters.

\section{Conclusions}

We have studied the GCS around the low-luminosity ($\rm M_V=-19.55$ mag) 
early-type galaxy NGC 7457. Several characteristics of the GCS are found 
typical for this type of galaxy: the specific frequency is found low with $\rm 
S_N \sim 2.7$ and the GC density profile is steep and follows the stellar light 
of the parent galaxy. The mean metallicity of $\rm <[Fe/H]>\ \simeq -1$ dex is 
compatible with the correlation existing between the metallicity of the GCSs 
and the luminosity of the parent galaxies.

The B-I and V-I color distributions are found unimodal but the B-I distribution 
is much wider than expected from a typical homogeneous population of GCs and 
consistent with what is expected for a bimodal distribution. The poor statistics 
(66 globular clusters in V-I and 41 in B-I) prevents from a detection of two 
peaks in the metallicity distribution of the GC system if they are closer than 1 
dex in [Fe/H] and/or if the two sub-populations have a very different number of 
objects.

Therefore our data are consistent with the presence of more than one population 
of GCs around NGC 7457. An alternative scenario would be a single GC population 
with a larger range in metallicity than found for homogeneous populations of 
GCs around the MW, nearby galaxies or bright ellipticals.

No significant population of metal-poor GCs similar to the halo GCs of the 
Galaxy is detected around NGC 7457. Several possibilities are discussed  to 
explain this absence of GCs with $\rm [Fe/H]\simeq -1.6 $ dex. 

Deeper photometry would be necessary to detect more GCs and to be able to 
identify two peaks in the metallicity distribution of the GCS. Another test of 
the hierarchical models of galaxy formation will be to estimate the age of the 
GCs since these models predict a rather recent merging for field early-type 
galaxies with $\rm z<1$. Such a determination requires spectroscopic 
observations of the GCs but will be difficult if the merging occurred more than 
5 Gyr ago since spectroscopic indices as $\rm H_{\beta}$ are not sensitive to 
higher ages. 

\begin{acknowledgements}

MK-P thanks the Alexander von Humboldt Foundation for its support through a 
Feodor Lynen Fellowship.

\end{acknowledgements}

\end{document}